\documentclass[12pt]{article}
\usepackage{latexsym}
\usepackage{epsfig,amssymb,euscript,slashed}
\usepackage{amsmath}
\usepackage{cite} 
\usepackage{array,calc,epsfig}
\usepackage{bbm}
\usepackage{fancybox}
\usepackage[linktocpage]{hyperref}

\def\be{\begin{equation}}
\def\ee{\end{equation}}
\def\bseq{\begin{subequations}}
\def\eseq{\end{subequations}}

\def\bea{\begin{eqnarray}}
\def\eea{\end{eqnarray}}

\def\bseq{\begin{subequations}}
\def\eseq{\end{subequations}}
\def\beq{\begin{equation}}
\def\eeq{\end{equation}}

\numberwithin{equation}{section} 
\usepackage[]{graphicx}

\def\sqr#1#2{{\vcenter{\vbox{\hrule height.#2pt
 \hbox{\vrule width.#2pt height#1pt \kern#1pt \vrule width.#2pt}\hrule
 height.#2pt}}}}

\def\slashchar#1{\setbox0=\hbox{$#1$}           
\dimen0=\wd0                                 
\setbox1=\hbox{/} \dimen1=\wd1               
\ifdim\dimen0>\dimen1                        
\rlap{\hbox to \dimen0{\hfil/\hfil}}      
#1                                        
\else                                        
\rlap{\hbox to \dimen1{\hfil$#1$\hfil}}   
/                                         
\fi}

\begin{document}
\font\cmss=cmss10 \font\cmsss=cmss10 at 7pt

\begin{flushright}{\scriptsize DFPD-14-TH-12 \\  \scriptsize QMUL-PH-14-13}
\end{flushright}
\hfill
\vspace{18pt}
\begin{center}
{\Large 
\textbf{Entanglement Entropy and D1-D5 geometries}
}
\end{center}

\vspace{8pt}
\begin{center}
{\textsl{ Stefano Giusto$^{\,a, b}$ and Rodolfo Russo$^{\,c}$}}

\vspace{1cm}

\textit{\small ${}^a$ Dipartimento di Fisica ed Astronomia ``Galileo Galilei",  Universit\`a di Padova,\\Via Marzolo 8, 35131 Padova, Italy} \\  \vspace{6pt}

\textit{\small ${}^b$ I.N.F.N. Sezione di Padova,
Via Marzolo 8, 35131 Padova, Italy}\\
\vspace{6pt}

\textit{\small ${}^c$ Centre for Research in String Theory, School of Physics and Astronomy\\
Queen Mary University of London,
Mile End Road, London, E1 4NS,
United Kingdom}\\
\vspace{6pt}

\end{center}

\vspace{12pt}

\begin{center}
\textbf{Abstract}
\end{center}

\vspace{4pt} {\small
\noindent 
In Conformal Field Theories with a gravitational AdS dual it is possible to calculate the entanglement entropy of a region $A$ holographically by using the Ryu-Takayanagi formula. In this work we consider systems that are in a pure state that is not the vacuum. We study in particular the 2D Conformal Field Theory dual to type IIB string theory on AdS$_3 \times S^3 \times T^4$ and focus on the $1/4$-BPS states described holographically by the 2-charge microstate geometries. We discuss a general prescription for the calculation of the entanglement entropy in these geometries that are asymptotically AdS$_3 \times S^3$. In particular we study analytically the perturbative expansion for a single, short interval: we show that the first non-trivial terms in this expansion are consistent with the expected CFT structure and with previous results on the vevs of chiral primary operators for the $1/4$-BPS configurations.}

\vspace{1cm}


\thispagestyle{empty}

\vfill
\vskip 5.mm
\hrule width 5.cm
\vskip 2.mm
{
\noindent  {\scriptsize e-mails:  {\tt stefano.giusto@pd.infn.it, r.russo@qmul.ac.uk} }
}

\setcounter{footnote}{0}
\setcounter{page}{0}

\newpage


\section{Introduction}

Entanglement entropies in quantum field theory have been at the centre of intense study in the last few years, in particular in the case of Conformal Field Theories (CFT) that admit a dual gravitational description. In $1+1$D CFTs, which will be the focus of this work, R\'enyi and von Neumann's Entanglement Entropies (EE) can be calculated in terms of correlators among local operators by using the replica trick~\cite{Calabrese:2004eu}. On the gravitational side, von Neumann's EE can be computed via the Ryu-Takayanagi (RT) formula~\cite{Ryu:2006bv} and the generalisation to R\'enyi's case was discussed in~\cite{Headrick:2010zt,Hartman:2013mia,Faulkner:2013yia}. A general argument explaining the RT formula has been recently given in \cite{Lewkowycz:2013nqa}. Most of the past work focused on density matrices $\rho_A$ obtained starting either from the $SL(2,\mathbb{C})$ invariant ground state (dual to AdS) or from the thermal state (dual to the BTZ black hole \cite{Caputa:2013lfa}) and tracing the degrees of freedom outside the space region $A$. When $A$ is an interval, the EE is given directly in terms of the central charge $c$ and does not depend on other details of the CFT. Things are more complicated if the space region $A$ is made of several disconnected intervals~\cite{Caraglio:2008pk,Furukawa:2008uk} and already the case of two disjoint intervals~\cite{Calabrese:2009ez, Calabrese:2010he} provides a good testing ground to study non-universal quantities.

In this work we focus on a different setup which also yields theory specific results: we study the EE for a density matrix obtained from a pure state $|s\rangle$ that is not the $SL(2,\mathbb{C})$ invariant vacuum. On the holographic side, it was first suggested in\footnote{We would like to thank B.~Vercnocke for bringing this paper to our attention and for an illuminating discussion on related issues~\cite{Dio:2014}.} \cite{Caputa:2013lfa} that the analysis of the EE in microstate geometries that are asymptotically AdS$_3$ represents a first step to understand microscopically the result for the extremal BTZ black hole. From the CFT point of view a similar problem has been recently analysed in~\cite{Bhattacharya:2012mi} and~\cite{Nozaki:2014hna,He:2014mwa,Caputa:2014vaa}. While the latter references focus on a time-dependent situation, we will focus, as in~\cite{Bhattacharya:2012mi}, on a density matrix obtained from an eigenstate $|s\rangle$ of the CFT Hamiltonian. This reference assumes that $|s\rangle$ is a small perturbation of the vacuum state. Since we aim to provide also a gravitational description of our analysis we focus on a $1+1$D CFT that has a well known string dual. The states we will be considering induce a macroscopic backreaction on the dual geometry and thus we need to consider the EE in a background that is not just AdS$_3$ plus a small perturbation. We argue that this requires a generalisation of the standard RT formula and check explicitly in some cases that the holographic results match the CFT expectations.

In particular we will focus on the superconformal field theory with $(4,4)$ supercharges and central charge $c=6 n_1 n_5$, whose dual gravitational description is given in terms of type IIB string theory compactified on $S^1\times T^4$ with $n_1$ D1-branes and $n_5$ D5-branes  wrapped on the compact space (the radius $R$ of the $S^1$ is much bigger than the string sized $T^4$ and all the branes wrap this circle). The gravitational description is appropriate for large charges $n_1,n_5\gg 1$ and for particular values of the moduli of the CFT. However it is convenient also to keep in mind a free field representation of the CFT with four bosonic and four fermionic fields whose target space is $(T^{4})^{n_1 n_5}/S_{n_1 n_5}$. This particular AdS/CFT duality has been thoroughly studied also because of its application to black hole physics in 5D: the Strominger-Vafa black hole~\cite{Strominger:1996sh} counts $1/8$-BPS states in this CFT and, in general, this setup can be used to address questions about the gravitational nature of each pure semiclassical state, a topic which is at the centre of the so-called fuzzball program\footnote{Recent reviews on the subject are~\cite{Skenderis:2008qn,Balasubramanian:2008da,Chowdhury:2010ct}, and a discussion of some general implications of this approach for the physics of black holes can be found in~\cite{Mathur:2012zp,Mathur:2012dxa,Bena:2013dka}.}~\cite{Mathur:2005zp}.

Our main goal is to study the EE for a single interval in the BPS states preserving $1/4$ of the $32$ supercharges of the type IIB theory. From the CFT point of view this means that we consider only the ground states in the Ramond-Ramond sector (i.e. the sector where the fermions have periodic boundary conditions). Of course these are eigenstates of the CFT Hamiltonian, with zero energy, so we are dealing with a stationary (but non-static) configuration. In particular we will focus on semiclassical states, which are dual to smooth geometries on the bulk side of the AdS/CFT correspondence. The general form of these solutions is known~\cite{Lunin:2001jy,Lunin:2002iz,Kanitscheider:2007wq} and we use it to compute holographically the EE of an interval. While the calculation can be set up in general, in order to give explicit results we focus on the limit where the size of the interval is small with respect to the $S^1$ where the CFT is defined (which coincides with the large $S^1$ in the string compactification). This limit allows for analytic calculations both on the gravity and the CFT sides, and, in the case of two intervals in the ground states, it was studied in~\cite{Calabrese:2009ez,Headrick:2010zt,Calabrese:2010he}. The first subleading term in this expansion is sufficient to show that there exists a state specific contribution beyond the leading universal result.

The outline of this paper is as follows. In Section~\ref{HEE} we first describe the general prescription for calculating the EE in a stationary geometry that is asymptotically AdS$~\times {\cal M}$, where ${\cal M}$ is a compact space. As the geometries we consider are generically non-static, our prescription generalizes the covariant Hubeny-Rangamani-Takayanagi (HRT) \cite{Hubeny:2007xt} formalism for the holographic computation of the EE. Then we focus on the $1/4$ BPS geometries of~\cite{Lunin:2001jy,Lunin:2002iz,Kanitscheider:2007wq}, which correspond to the Ramond-Ramond ground states of the dual CFT. The result for the EE of a single interval is given in terms of an integral which includes the compact space. In Section~\ref{sil} we discuss in detail the short interval limit up to the first non-universal terms. In Section~\ref{ccft} we re-interpret the gravity result in terms of the underlying CFT. The quantity under analysis is non-protected and so it is not possible to use directly the free orbifold description. However, the short interval expansion can be naturally written in a way to separate the contributions of the BPS operators, which survive in the strong coupling limit where the gravity approximation is valid,  from the others. We show that, by focusing on the protected operators, one can recover the gravity result discussed previously. To the best of our knowledge, this result represents the first non-trivial check of the RT formula (or more precisely of its 6D extension) in a situation where the EE has also non-universal contributions.

\section{Holographic entanglement entropy}\label{HEE}

In theories that admit a holographic dual, the EE can be computed via the Ryu-Takayanagi (RT) formula\cite{Ryu:2006bv}. In its simplest form, the formula applies to theories whose gravity dual is classical Einstein gravity\footnote{The generalisation to holographic theories with higher curvature corrections has been worked out in~\cite{Dong:2013qoa,Camps:2013zua}, see also \cite{Bhattacharyya:2014yga}.} (eventually  plus matter) and to states dual to static classical spacetimes that tend asymptotically to AdS$_{d+1}$. For $d=2$, the EE of a one-dimensional spatial region $A$ is given by
\be\label{RT}
S_A = \frac{\mathrm{area} (\gamma_A)}{4 G_N}\,,
\ee
where $\gamma_A$ is the curve of minimal length homologous to $A$, in the space slice of the bulk containing $A$, and $G_N$ is the Newton's constant of the 3-dimensional theory.

We aim to apply the RT formalism to compute entanglement entropies in states of the D1-D5 black hole. These states can be identified with the RR ground states of a 2D CFT, that we will denote as the D1-D5 CFT (for a review see \cite{Avery:2010qw}). The gravitational duals of these states
are described semiclassically by 10D supergravity solutions that, in the decoupling limit, are asymptotically AdS$_3\times S^3\times T^4$. As the $T^4$ is taken to have string size, the geometries are smeared on the $T^4$ and can be equivalently described by 6D solutions, with Einstein metric $ds^2_6$. Generic microstates depend however non-trivially on the $S^3$ directions, and there is no canonical way to reduce them to 3D asymptotically AdS$_3$ solutions. We thus need a generalization of the RT formula (\ref{RT}), that applies to 6D spacetimes asymptotic to AdS$_3\times S^3$. Given a 1D spatial region $A$, we propose that its EE in a D1-D5 microstate is given by
\be\label{RTbis}
S_A = \frac{\mathrm{area} (\Gamma_A)}{4 G'_N}\,,
\ee
where $\Gamma_A$ is the 4D minimal-area surface of the 6D geometry at constant time that at the AdS$_3$ boundary reduces to $\partial A \times S^3$ and in the bulk has the product structure defined below; $G'_N$ is the 6D Newton's constant. In order to provide a precise definition of the class of 4D manifolds to which $\Gamma_A$ belongs, one needs to give meaning to the split of the 6D space into an AdS$_3$ and an $S^3$ part or in other words to introduce an almost product structure. While this split can be unambiguously defined at the boundary of the space, where the geometry reduces to AdS$_3\times S^3$, there are  various inequivalent ways to extend it in the interior of the bulk\footnote{We are very grateful to J. Simon for this observation and also for drawing our attention to the relevance of the covariant HRT prescription in the context of microstate geometries, as discussed at the end of this section.}. An almost product structure can be defined by choosing a system of coordinates $x^I = (x^\mu, x^\alpha)$ (with $I=1,\ldots 6$, $\mu=1,\ldots,3$, $\alpha=1,\ldots,3$), where, at the boundary, $x^\mu$ and $x^\alpha$ are coordinates in AdS$_3$ and $S^3$. These coordinates are extended in the bulk in such a way that the 6D Einstein metric $G_{IJ}$  satisfies the de Donder-Lorentz gauge\footnote{The de Donder-Lorentz gauge is the one commonly employed when reducing on compact spaces \cite{Kim:1985ez}, and it seems natural in the AdS/CFT context, because it was shown in \cite{Skenderis:2006uy} that it reproduces the results of a gauge-invariant KK reduction procedure.}
\be\label{deDonder}
\nabla^\alpha \hat G_{\alpha\beta} = \nabla^\alpha G_{\alpha\mu}=0\,,
\ee
where covariant derivatives are defined with respect to the round $S^3$ metric and $\hat G_{\alpha\beta}$ is the traceless part of $G_{\alpha\beta}$. Using these coordinates, the 6D Einstein metric can be written in the form
\be\label{6Deinstein}
ds^2_6 = G_{IJ} dx^I dx^J = g_{\mu\nu} dx^\mu dx^\nu + G_{\alpha\beta} (dx^\alpha + A^\alpha_\mu dx^\mu)(dx^\beta + A^\beta_\nu dx^\nu)\,,
\ee
which defines the split of the 6D metric into a deformed AdS$_3$ and $S^3$ parts indicated as $g_{\mu\nu}$ and $G_{\alpha\beta}$ respectively. As usual in KK reductions, this split is invariant under reparametrizations of the compact space ($x^\alpha \to x^\alpha (x^\mu,x^\beta)$), but is not invariant under $x^\alpha$-dependent changes of the coordinates $x^\mu$ ($x^\mu \to x^\mu(x^\nu,x^\alpha)$); it is precisely this arbitrariness that is fixed by the gauge condition (\ref{deDonder}).
So the prescription we propose is to minimise the functional~\eqref{RTbis} over the class of 4-manifolds that are invariant under the almost product structure induced by the coordinate split~\eqref{6Deinstein}. These manifolds can be parametrized as $x^I(\lambda,x^\alpha) = (x^\mu(\lambda),x^\alpha)$, where $x^0(\lambda)=\mathrm{const.}$ when $\Gamma_A$ lies in a constant time slice. The metric induced on the 4-manifold $\Gamma_A$ is
\be
ds^2_* = g_{\mu\nu} \dot{x}^\mu \dot{x}^\nu d\lambda^2 + G_{\alpha\beta} (dx^\alpha + A^\alpha_\mu\, \dot{x}^\mu d\lambda)(dx^\beta + A^\beta_\nu \,\dot{x}^\nu d\lambda)\,,
\ee
its determinant is
\be
\mathrm{det}(g_*) = g_{\mu\nu} \dot{x}^\mu \dot{x}^\nu\,\mathrm{det} (G_{\alpha\beta})\,,
\ee
and the area of the 4-manifold is
\be\label{area}
\mathrm{area} (\Gamma_A) = \int \!d\lambda\, d^3x^\alpha\,\sqrt{\mathrm{det} (G_{\alpha\beta})}\,\sqrt{g_{\mu\nu} \dot{x}^\mu \dot{x}^\nu}\equiv \int \!d\lambda\, d^3x^\alpha\,\sqrt{g^E_{\mu\nu} \dot{x}^\mu \dot{x}^\nu} \,,
\ee
where we have defined
\be\label{ads3einstein}
g^E_{\mu\nu} = g_{\mu\nu}\,   \mathrm{det}(G_{\alpha\beta})\,.
\ee
For generic microstates $g^E_{\mu\nu}$ depends non-trivially on the $S^3$ coordinates $x^\alpha$. When instead  $g^E_{\mu\nu}$ in~\eqref{ads3einstein} is independent of $x^\alpha$, it is just the Einstein metric of the 3D theory reduced on $S^3$, and the prescription~(\ref{RTbis}) reduces to the RT formula~\eqref{RT} for the asymptotically AdS$_3$ metric $g^E_{\mu\nu}$.

Generic microstates are, moreover, associated with stationary but non-static geometries. It was shown in \cite{Hubeny:2007xt} that for non-static geometries the RT prescription has to be generalized by relaxing the constraint that the class of manifolds over which one minimizes the area functional lie in a constant time slice. In this more general setting, minimal surfaces might no longer exist, and one should look instead for extremal surfaces. We will denote this covariant generalization of the RT prescription as the HRT prescription. The HRT formalism can be generalized to space times asymptotic to AdS$_3\times S^3$ along the same lines outlined above: the covariant 6D prescription is to find the extrema of the area functional (\ref{area}) over manifolds $\Gamma_A$ that are invariant under the almost product structure previously defined, without imposing any restriction on $x^0(\lambda)$ in the bulk. 

\subsection{Solution of the geodesic problem for a single interval}
\label{sec:geodesic}
Let us now work out the equations satisfied by extremal surfaces in a general non-static geometry.
For the purpose of extremizing the area functional (\ref{area}) with respect to the functions $x^\mu(\lambda)$, the $S^3$ coordinates $x^\alpha$ play the role of external parameters, and explicit dependence on $x^\alpha$ will be suppressed in the following: it is understood that everything is computed at some fixed value of $x^\alpha$, over which one integrates at the end. As usual, it is convenient to parametrize $x^\mu(\lambda)$ in terms of the ``proper time'' parameter $\tau$, satisfying
\be\label{propertime}
g^E_{\mu\nu} \dot{x}^\mu\dot{x}^\nu=1\,.
\ee
In this parametrization, extrema of the area functional satisfy 
\be\label{geodesics}
\frac{d}{d\tau} (g^E_{\mu\nu} \dot{x}^\mu)=\frac{1}{2} \partial_\nu g^E_{\mu\lambda} \dot{x}^\mu \dot{x}^\lambda\,.
\ee
Two-charge microstate geometries do not depend on the angular coordinate of AdS$_3$, which is identified with the spatial coordinate of the CFT and will be denoted by $y$, but only on the AdS$_3$ radial coordinate $r$ (apart from $x^\alpha$). Moreover we will assume for simplicity the gauge $g^E_{rt}=g^E_{ry}=0$;  as we will see it is straightforward to satisfy this condition at the leading order in the large $r$ expansion. The relevant metric components are then $g^E_{rr}(r)$ and $g^E_{mn}(r)$ where we denote by $m,n$ indices that take values $t,y$. 
The components $\nu=n=t,y$ of the extremality equations (\ref{geodesics}) give
\be\label{ytau}
\frac{d}{d\tau} (g^E_{m n}(r) \dot{x}^m) = 0 \quad \Rightarrow\quad \dot{x}^m = g^{m n}_E(r) \kappa_n\,,
\ee
where $g^{mn}_E$ is the inverse of $g^E_{m n}$ and $\kappa_n$ are constants (that might depend on $x^\alpha$).
The constraint (\ref{propertime}) implies
\be\label{rtau}
g^E_{rr}(r) \dot{r}^2 + g^E_{mn}(r)\, \dot{x}^m \dot{x}^n=1 \quad \Rightarrow\quad \dot{r}^2 = \frac{1-g^{mn}_E(r)\, \kappa_m\, \kappa_n}{g^E_{rr}(r)} \,.
\ee
Eqs. (\ref{ytau}), (\ref{rtau}) determine $x^\mu(\tau)$ after specifying the boundary conditions which depend on the choice of the spatial region $A$.  We will restrict to spatial regions made of a single interval of length $l$ at $t=\bar t$. The end-points of the curve $x^\mu(\tau)$ at the boundary of AdS$_3$ ($r\to \infty$) have to coincide with the boundaries of the interval $A$. Since the area of a 4-manifold that extends all the way to $r=\infty$ diverges, to obtain a finite result it is necessary to introduce an IR cut-off $r_0$ and replace the AdS$_3$ boundary with the surface $r=r_0$ . This explains the choice of the following boundary conditions:
\be\label{eq:bc}
r(\tau_1)=r_0\,,\quad t(\tau_1) = \bar t\,,\quad y(\tau_1)=0\,;\quad r(\tau_2)=r_0\,,\quad t(\tau_2) = \bar t\,,\quad y(\tau_2)=l\,.
\ee
Then
\be
\label{eqct}
0 = \int_{\tau_1}^{\tau_2} \dot{t}\, d\tau = 2 \int_{r^*}^{r_0} \frac{\dot{t}}{\dot{r}}\, dr = 2 \kappa_m  \int_{r^*}^{r_0} \! dr\,g^{tm}_E(r)\sqrt{\frac{g^E_{rr}(r)}{1-g^{np}_E(r)\, \kappa_n\, \kappa_p}}\,,
\ee
\be\label{eqc}
l = \int_{\tau_1}^{\tau_2} \dot{y}\, d\tau = 2 \int_{r^*}^{r_0} \frac{\dot{y}}{\dot{r}}\, dr =  2 \kappa_m  \int_{r^*}^{r_0} \! dr\,g^{ym}_E(r)\sqrt{\frac{g^E_{rr}(r)}{1-g^{np}_E(r)\, \kappa_n\, \kappa_p}}\,,
\ee
where the turning point $r^*$ is the largest solution of  
\be\label{eq:turpoi}
g^{mn}_E(r^*)\, \kappa_m\, \kappa_n=1\,.
\ee
Inverting Eqs.~(\ref{eqct}-\ref{eqc}) determines the parameters $\kappa_m$ in terms of the interval length $l$. These values of $\kappa_m$ can then be replaced in the expression for the area of the minimal submanifold $\Gamma_A$
\be\label{areamA}
\mathrm{area}(\Gamma_A) =  \int d^3 x^\alpha \int_{\tau_1}^{\tau_2} d\tau = 2 \int d^3 x^\alpha \int_{r^*}^{r_0} \frac{1}{\dot{r}}\, dr = 2\int d^3 x^\alpha \int_{r^*}^{r_0} \!dr\,\sqrt{\frac{g^E_{rr}(r)}{1-g^{mn}_E(r)\, \kappa_m\, \kappa_n}}\,.
\ee
According to (\ref{RTbis}), the EE of the interval $A$ is then
\be\label{SA}
S_A= \frac{\mathrm{area}(\Gamma_A) }{4 G'_N}= \frac{c}{6}\frac{\mathrm{area}(\Gamma_A)}{\mathrm{vol}(S^3_b)R_{AdS}} = n_1 n_5\, \frac{\mathrm{area}(\Gamma_A)}{\mathrm{vol}(S^3_b) R_{AdS}}\,,
\ee
where $\mathrm{vol}(S^3_b)$ is the volume of the 3D sphere at the boundary of AdS. We also used
\be\label{eq:Gn1n5}
G'_N = \mathrm{vol}(S^3_b) G_N \,,\quad c =\frac{3}{2}\frac{R_{AdS}}{G_N} = 6 \,n_1 n_5\,,
\ee
with $n_1$, $n_5$ the numbers of D1 and D5 branes and $R_{AdS}$ the radius of AdS.

\section{Entanglement entropy in D1-D5 states for small \texorpdfstring{$l$}{l}}\label{sil}

The geometry of generic D1-D5 states has been constructed in \cite{Kanitscheider:2007wq}, whose conventions we will follow here. We will restrict for simplicity to the class of states that are invariant under rotations in the internal $T^4$ directions (for which $\mathcal{A}^{\alpha-}=0$, in the notation of~\cite{Kanitscheider:2007wq}). The 6D Einstein metric of these states is
\be\label{generic2charge}
ds^2_6=f^{-1}[-(dt-A)^2 + (dy-B)^2] + f \,dx^i dx^i \,,
\ee
where 
\be\label{eq:fdef}
f \equiv (f_1 f_5 - \mathcal{A}^2)^{1/2}\,,
\ee
$x^i$ ($i=1,\dots,4$) are coordinates in $\mathbb{R}^4$ and $A\equiv A_i \,dx^i$, $B\equiv B_i \,dx^i$ are 1-forms on $\mathbb{R}^4$ that satisfy $dB= - *_4 dA$ .  $f_1$, $f_5$, $\mathcal{A}$, $A_i$ are harmonic functions on $\mathbb{R}^4$ whose explicit expressions are, for instance, given in Eqs.~(2.12) and~(2.5) of~\cite{Kanitscheider:2007wq}.

The simplest microstates are the ones with maximal or minimal values of the $SU(2)_L\times SU(2)_R$ R-charges: $j = \pm n_1 n_5/2$, $\bar j = \pm n_1 n_5/2$. For example the geometry of the state with $j=\bar j = n_1 n_5/2$ is, in the decoupling limit,
\be\label{circularLM}
ds^2_6 = - \frac{\hat r^2+a^2}{\sqrt{Q_1 Q_5}}dt^2 + \frac{\hat r^2}{\sqrt{Q_1 Q_5}} dy^2 + \sqrt{Q_1 Q_5} \frac{d\hat r^2}{\hat r^2+a^2} + \sqrt{Q_1 Q_5}\, (d\hat \theta^2 + \cos^2\hat \theta d\hat \psi^2 + \sin^2\hat \theta d\hat \phi^2)\,,
\ee
with
\be\label{spectralflow}
\hat \psi= \psi - \frac{y}{R}\,,\quad \hat \phi = \phi - \frac{t}{R}\,.
\ee
The parameter $a$ is related to the D1 and D5 charges, $Q_1$ and $Q_5$, and the radius $R$ of the $S^1$ direction $y$, by
\be\label{radius}
a=\frac{\sqrt{Q_1 Q_5}}{R}\,.
\ee
The coordinates $(\hat r, \hat\theta)$ are mapped to polar coordinates of $\mathbb{R}^4$ $(r,\theta)$ by
\be
r^2 = \hat r^2 + a^2 \sin^2\hat \theta\,,\quad \cos^2\theta =\frac{ \hat r^2 \cos^2\hat \theta}{\hat r^2 + a^2 \sin^2\hat \theta}\,.
\ee
It is immediate to check from (\ref{circularLM}) that these coordinates satisfy the de Donder-Lorentz gauge (\ref{deDonder}).
Hence, in $\hat r, \hat \theta, \hat \psi, \hat \phi$ coordinates, it becomes explicit that the 6D geometry of this particular microstate is simply AdS$_3\times S^3$, and the 3D geometry $g^E_{\mu\nu}$ reduced on $S^3$ is just global AdS$_3$. According to the recipe (\ref{RTbis}), the EE of the interval $A=[0,l]$ computed in this state is the same as the one in the $SL(2,\mathbb{C})$-invariant vacuum:
\be
S_A= 2 n_1 n_5 \log \Bigl[\frac{2 r_0}{a} \sin \Bigl(\frac{l}{2 R}\Bigr)\Bigr]\,.
\ee

The metrics for generic microstates are too complicated to analytically carry out the holographic EE computation exactly. A limit which is amenable to analytic computations, both on the gravity and on the CFT side, is the short interval regime, in which $l$ is much smaller than the $S^1$ radius $R$. In this limit the extremal submanifold $\Gamma_A$ only probes the region of the geometry close to the boundary: hence only the large $r$ expansion of the geometry (\ref{generic2charge}) is relevant in this approximation. We will consider just the first non-trivial correction in the $l$ expansion, and for this purpose one can approximate the metric coefficients as
\begin{align}\label{eq:asymp}
f_1\approx & \frac{Q_1}{r^2}\Bigl(1+\frac{f^1_{1i}}{r} Y^i_1 +\frac{f^1_{2I}}{r^2} Y^I_2 \Bigr)\,,\,\, f_5\approx\frac{Q_5}{r^2}\Bigl(1+\frac{f^5_{1i}}{r} Y^i_1 +\frac{f^5_{2I}}{r^2} Y^I_2 \Bigr)\,,\,\,
\mathcal{A}\approx\frac{\sqrt{Q_1 Q_5}\,\mathcal{A}_{1i}}{r^3} \,Y^i_1\,,
\nonumber \\
A \approx & \frac{\sqrt{Q_1 Q_5}}{r^2} \, (a_{\alpha+} Y^{\alpha+}_1 + a_{\alpha-} Y^{\alpha-}_1)\,,\,\, B \approx\frac{\sqrt{Q_1 Q_5}}{r^2} \, (a_{\alpha+} Y^{\alpha+}_1 - a_{\alpha-} Y^{\alpha-}_1)\,.
\end{align}
Here $f^{1}_{kI}$, $f^{5}_{kI}$, $\mathcal{A}_{1i}$, $a_{\alpha\pm}$ are constants that can be computed once a specific 2-charge microstate geometry is chosen. In the small $l$ expansion we are considering, we will only keep terms up to second order in $f^{1,5}_{1i}$,  $\mathcal{A}_{1i}$ and $a_{\alpha\pm}$ and up to first order in  $f^{1,5}_{2I}$.  It is always possible to pick coordinates in such a way that
\be
f^{1}_{1i}+ f^{5}_{1i}=0\,,
\ee
and we will take advantage of this gauge choice in the following. $Y^I_k$ are scalar spherical harmonics of degree $k$ on $S^3$. We will need in particular the harmonics of degree 1: the scalar $Y^i_1$, with $i=1,\ldots,4$, and the vector harmonics $Y^{\alpha\pm}_1$, with $\alpha=1,2,3$, are 
\be\label{Ydeg1}
Y^i_1 = 2 \frac{x^i}{r}~,\quad Y^{\alpha+}_1= \frac{ \eta^\alpha_{ij}\,dx^i x^j}{r^2}\,,\quad Y^{\alpha-}_1= \frac{\bar\eta^\alpha_{ij} \,dx^i x^j}{r^2}\,,
\ee
where $\eta^\alpha_{ij} = \delta_{\alpha i} \,\delta_{4j}- \delta_{\alpha j} \,\delta_{4i} + \epsilon_{\alpha i j 4}$, $\bar \eta^\alpha_{ij}= \delta_{\alpha i} \,\delta_{4j}- \delta_{\alpha j} \,\delta_{4i} - \epsilon_{\alpha i j 4}$  are the standard 't Hooft symbols. One can use either $Y^{\alpha+}_1$ or $Y^{\alpha-}_1$ to form a basis of 1-forms on $S^3$, and moreover the round $S^3$ metric can be 
written as $ds^2_3= \sum_\alpha Y^{\alpha+}_1 \otimes Y^{\alpha+}_1=\sum_\alpha Y^{\alpha-}_1 \otimes Y^{\alpha-}_1$. In order to rewrite the metric in the form (\ref{6Deinstein}), suitable to perform the reduction on $S^3$, it is convenient to express the 1-forms in one of the two basis, let us say $Y^{\alpha+}_1$. Hence we write
\be
Y^{\alpha-}_1=R^\alpha_\beta\, Y^{\beta+}_1\,,
\ee
where the coefficients $R^\alpha_\beta$ depend on the $S^3$ coordinates, and
\be
A \approx\frac{\sqrt{Q_1 Q_5}}{r^2} \, (a_{\alpha+}  + \tilde a_{\alpha-})Y^{\alpha+}_1\,,\,\, B \approx\frac{\sqrt{Q_1 Q_5}}{r^2} \, (a_{\alpha+}  - \tilde a_{\alpha-})Y^{\alpha+}_1\,,
\ee
with
\be
\tilde a_{\alpha -} = R_\alpha^\beta \, a_{\beta -}\,. 
\ee
The scalar and vector spherical harmonics in~\eqref{Ydeg1} satisfy
\begin{align}
& (Y^{\alpha+}_1)_\gamma (Y^{\beta+}_1)^\gamma = \delta^{\alpha\beta}\,,\quad 
(Y^{\alpha-}_1)_\gamma (Y^{\beta-}_1)^\gamma = \delta^{\alpha\beta}\,,
\label{vectorharm}
\\ \nonumber
& \frac{1}{2\pi^2}\int_{S^3}d\Omega_3 \,Y_1^i Y_1^j = \delta^{ij}\,,\quad \int_{S^3}d\Omega_3 \,Y_k^I=0\,,\quad \int_{S^3} d\Omega_3 (Y^{\alpha+}_1)_\gamma (Y^{\beta-}_1)^\gamma = 0\,,
\end{align}
where the contraction over the $S^3$ indices $\gamma$ and the volume form $d\Omega_3$ are the ones associated with the round $S^3$ metric. This implies
\be
R^\alpha_\gamma\,R^\beta_\gamma = \delta^{\alpha\beta}\,,\quad \int_{S^3} R^\alpha_\beta=0\,.
\ee
The system of coordinates used in Eqs. (\ref{eq:asymp}) does not satisfy the de Donder-Lorentz gauge conditions (\ref{deDonder}) at the required order in the perturbative expansion. Before extracting the 3D metrics $g_{\mu\nu}$ and $G_{\alpha\beta}$ from (\ref{6Deinstein}), one should thus change to coordinates satisfying the gauge (\ref{deDonder}). We have checked that, at our perturbative order, this procedure generates corrections to the $g_{\mu\nu}^E$ computed in the coordinates of  (\ref{eq:asymp}) that are linear in the scalar harmonics of degree 2; hence these corrections vanish when integrated over $S^3$, thanks to the properties of spherical harmonics (\ref{vectorharm}). For simplicity of exposition, we will thus continue working with the coordinates of (\ref{eq:asymp}).

At the required order in $1/r$, the generic 2-charge metric (\ref{generic2charge}) can be recast in the form
\begin{align}
\nonumber
ds^2_6 &\approx f^{-1}\!\Bigl[-\Bigl(1+\frac{(a_{\alpha+}  + \tilde a_{\alpha-})^2}{r^2}\Bigr) dt^2+\Bigl(1-\frac{(a_{\alpha+}  - \tilde a_{\alpha-})^2}{r^2}\Bigr) dy^2-2 \frac{(a_{\alpha+})^2 - (a_{\alpha-})^2}{r^2}dt dy\Bigr]\\
&+f (dr^2 + r^2\,  \hat Y^{\alpha+}_1\, \hat Y^{\alpha+}_1)-2\frac{\sqrt{Q_1 Q_5}}{r^2}\,a_{\alpha+}\tilde a_{\beta-} \, \hat Y^{\alpha+}_1\, \hat Y^{\beta +}_1\,,
\label{generic6D}
\end{align}
with
\be
\hat Y^{\alpha+}_1 = Y^{\alpha+}_1 + \frac{a_{\alpha+}+\tilde a_{\alpha-}}{\sqrt{Q_1 Q_5}} dt - \frac{a_{\alpha+}-\tilde a_{\alpha-}}{\sqrt{Q_1 Q_5}} dy\,,
\ee
and 
\be
f \approx \frac{\sqrt{Q_1 Q_5}}{r^2}\Bigl(1-\frac{f^1_{1i} f^1_{1j}+\mathcal{A}_{1i} \mathcal{A}_{1j}}{2 r^2} Y^i_1 Y^j_1 + \frac{f^1_{2I}+f^5_{2I}}{2 r^2} Y^I_2\Bigr)\,.
\ee
The determinant of the $S^3$ metric $G_{\alpha\beta}$ at order $1/r^2$ can be read off from (\ref{generic6D})
\be
\mathrm{det}(G_{\alpha\beta})\approx(f r^2)^3 \det{G_{S^3}} \Bigl(1-2\, \frac{a_{\alpha+}\,a_{\beta-}}{r^2} \,(Y^{\alpha+}_1)_\gamma (Y^{\beta-}_1)^\gamma\Bigr)\,,
\ee
where $ \det{G_{S^3}}$ is the determinant of the metric for a round 3-sphere of unit radius. In the same approximation the AdS$_3$ metric defined in (\ref{ads3einstein}) is 
\begin{align}
\nonumber
ds^2_E&\approx (f r^2)^3 \sin^2\theta \cos^2\theta \Bigl(1-2\, \frac{a_{\alpha+}\,a_{\beta-}}{r^2} \,(Y^{\alpha+}_1)_\gamma (Y^{\beta-}_1)^\gamma\Bigr)\,\times\\
\nonumber
&\times \Bigl\{f^{-1}\!\Bigl[-\Bigl(1+\frac{(a_{\alpha+}  + \tilde a_{\alpha-})^2}{r^2}\Bigr) dt^2+\Bigl(1-\frac{(a_{\alpha+}  - \tilde a_{\alpha-})^2}{r^2}\Bigr) dy^2\\
\nonumber
&\qquad-2 \frac{(a_{\alpha+})^2 - (a_{\alpha-})^2}{r^2}dt dy\Bigr]+f dr^2\Bigr\}\\
\nonumber
& \equiv (Q_1 Q_5)^{3/2}\sin^2\theta\cos^2\theta\Bigl[ \frac{r^2}{\sqrt{Q_1 Q_5}} [-(1+\delta g^E_{tt}) dt^2 + (1+\delta g^E_{yy}) dy^2 + 2\, \delta g^E_{ty}\, dt dy]\\ \label{generic3D}
&\quad +\frac{\sqrt{Q_1 Q_5}}{r^2}(1+\delta g^E_{rr}) dr^2 \Bigr]\,.
\end{align}
The metric $ds^2_E$ in general depends on the $S^3$ coordinates, but for the purpose of computing the area functional $\mathrm{area}(\Gamma_A)$ one can introduce a ``reduced'' AdS$_3$ metric $d\hat s^2_E$, integrated over $S^3$, such that
\be
\frac{\mathrm{area}(\Gamma_A)}{4 G'_N} = \frac{\int \!d\lambda\, d^3x^\alpha\,\sqrt{g^E_{\mu\nu} \dot{x}^\mu \dot{x}^\nu}}{4 G'_N} = \frac{\int \!d\lambda\,\sqrt{\hat g^E_{\mu\nu} \dot{x}^\mu \dot{x}^\nu}}{4 G_N}\,.
\ee
This reduced metric is given by
\be
d\hat s^2_E = \frac{r^2}{\sqrt{Q_1 Q_5}} [-(1+\delta \hat g^E_{tt}) dt^2 + (1+\delta \hat g^E_{yy}) dy^2 + 2\, \delta \hat g^E_{ty}\, dt dy] +\frac{\sqrt{Q_1 Q_5}}{r^2}(1+\delta \hat g^E_{rr}) dr^2\,,
\ee
where
\be
\delta \hat g^E_{\mu\nu} = \frac{1}{2\pi^2}\int_{S^3} d\Omega_3\, \delta g^E_{\mu\nu}\,.
\ee
Comparing with (\ref{generic3D}), and using (\ref{vectorharm}) one finds
\begin{subequations}
\be
\delta \hat g^E_{tt}=\frac{a_+^2+a_-^2-f_1^2-\mathcal{A}_1^2}{r^2}\,,\quad \delta \hat g^E_{yy}=-\frac{a_+^2+a_-^2+f_1^2+\mathcal{A}_1^2}{r^2}\,,
\ee
\be
\delta \hat g^E_{ty}=-\frac{a_+^2-a_-^2}{r^2}\,,\quad \delta \hat g^E_{rr}=-2\,\frac{f_1^2+\mathcal{A}_1^2}{r^2}\,,
\ee
\end{subequations}
where we introduced the condensed notation
\be
a_{\pm}^2 \equiv a_{\alpha\pm} a_{\alpha\pm}\,,\quad f_1^2 \equiv f^1_{1i} f^1_{1i}\,,\quad \mathcal{A}_1^2 \equiv \mathcal{A}_{1i} \mathcal{A}_{1i}\,.
\ee

Let us now apply the formalism of Section~\ref{sec:geodesic} and determine the extremal curves for the reduced metric $\hat g^E_{\mu\nu}$. Note that $\hat g^E_{ty}$ is non-trivial and thus we should use the covariant prescription, without restricting to constant $x^0(\lambda)$. The EE, however, is invariant under changes of the orientation of the space ($y\to -y$) and hence it will depend at least quadratically on $\hat g^E_{ty}$\footnote{This conclusion can also be verified directly from the equations of Section~\ref{sec:geodesic}: it follows from (\ref{eqct}) that $\kappa_t$ is of oder $\hat g^E_{ty}$ and from (\ref{eqc}) and (\ref{areamA}) one sees that the EE receives contributions that are either proportional to $\kappa_t^2$ or to $\hat g^E_{ty} \kappa_t$.}. As in our case $\hat g^E_{ty}$ is already quadratic in $a_{\alpha\pm}$, its contributions to the EE will be quartic in $a_{\alpha\pm}$, and will be discarded at our perturbative order. We can thus simplify the computation and take $x^0(\lambda)=\mathrm{const.}$ and $\kappa_t=0$. Then eq.~(\ref{eqc}) reads  (at our level of approximation)
\be\label{eqhatc}
\begin{aligned}
l &\approx 2\, \kappa_y\, (Q_1 Q_5)^{3/4} \int_{\hat \kappa}^\infty\frac{dr}{r^2\sqrt{r^2-\hat \kappa^2}}\Bigl(1+\frac{a_+^2+a_-^2-f_1^2 - \mathcal{A}_1^2}{2 r^2}\Bigr)\\
& = 2\, \kappa_y\, (Q_1 Q_5)^{3/4} \Bigl(\frac{1}{\hat \kappa^2} +\frac{a_+^2+a_-^2-f_1^2 - \mathcal{A}_1^2}{3 \hat \kappa^4}\Bigr)\\
& \approx \frac{2\,\sqrt{Q_1 Q_5}}{\hat \kappa}\Bigl(1-\frac{a_+^2+a_-^2 + 5(f_1^2 + \mathcal{A}_1^2)}{6\hat \kappa^2}\Bigr)  \,,
\end{aligned}
\ee
where in the last step we expanded for small $l$ and used the following expression for the turning point
\be
(r^*)^2\equiv \hat \kappa^2  = \sqrt{Q_1 Q_5}\, \kappa_y^2 + a_+^2 + a_-^2 + f_1^2 + \mathcal{A}_1^2\,.
\ee
Note that we sent $r_0\to \infty$ because the above integral is convergent. Eq.~(\ref{eqhatc}) should be inverted to express $\hat \kappa$ in terms of $l$:
\be
\hat \kappa \approx \frac{2\sqrt{Q_1 Q_5}}{l}\Bigl(1-\frac{a_+^2 + a_-^2 +5(f_1^2 + \mathcal{A}_1^2)}{24\,Q_1 Q_5}\, l^2\Bigr)\,.
\ee
We can now use Eqs.~(\ref{areamA}) and~(\ref{SA}), and the fact that $R_{AdS}= (Q_1 Q_5)^{1/4}$, to compute the EE for the interval $A=[0,l]$ in a generic 2-charge state at order $l^2$:
\be\label{gravityresult}
\begin{aligned}
S_A&\approx 2\, n_1 n_5\int_{\hat \kappa}^{r_0}\frac{dr}{\sqrt{r^2-\hat \kappa^2}}\Bigl(1-\frac{a_+^2+a_-^2+3(f_1^2+\mathcal{A}_1^2)}{2 r^2}\Bigr)\\
&=2\,n_1 n_5 \Bigl(\log \frac{r_0 (1+\sqrt{1-\hat \kappa/r_0})}{\hat \kappa} - \frac{a_+^2+a_-^2+3(f_1^2+\mathcal{A}_1^2)}{2\hat \kappa^2}\Bigr)\\
&\approx 2\,n_1 n_5 \Bigl(\log \frac{r_0 \,l}{\sqrt{Q_1 Q_5}} - \frac{a_+^2 + a_-^2 + 2 (f_1^2 + \mathcal{A}_1^2)}{12 \,Q_1 Q_5}\,l^2\Bigr)\,.
\end{aligned}
\ee

\section{Comparing with the CFT expectation}\label{ccft}

In this section we show how to interpret the result in Eq.~\eqref{gravityresult} from the CFT point of view. First we need to introduce a density matrix $\rho_A^{(s)}$ that is obtained by starting from a pure state $|s\rangle$ of the CFT and by tracing over the degrees of freedom in the complement of region $A$. We restrict ourselves to the case where $A$ is a single interval and $|s\rangle$ is an eigenstate of the CFT Hamiltonian, so the time evolution of the problem is trivial. Even for these simple situations, the EE in general depends on all details of the CFT. Thus, in order to have an analytic approach and match the supergravity result, we focus on the short interval limit as discussed above. 

As usual~\cite{Calabrese:2004eu}, we start by considering $n$ independent copies of the original CFT and then insert at the endpoints of the interval $A$ twist fields ${\cal T}_{\pm n}$ that introduce a monodromy which identifies two consecutive sheets. For instance if $T_j(z)$ is the (holomorphic part of the) stress energy tensor defined on the $j^{\rm th}$ copy, then $T_j \to T_{j\pm 1}$ when it goes around the operator ${\cal T}_{\pm n}$. The same monodromy holds also for the anti-holomorphic fields. Properties and correlators of twist fields have been extensively discussed in several contexts; for a discussion inspired by AdS/CFT see \cite{Jevicki:1998bm,Lunin:2000yv,Lunin:2001pw}. Even the simplest correlators in presence of twist fields are defined on a complicated worldsheet that is obtained by gluing at the positions of the ${\cal T}$'s the different sheets where each copy of the CFT is defined. In general the EE of the interval $A=[0,l]$ in the state $|s\rangle$ is given by\footnote{R\'enyi entropies $S^{(s) R}_n$ are related to $S^{(s)}_n$ as 
\be
S^{(s) R}_n = \frac{1}{1-n} \log S^{(s)}_n\,.
\ee}
\be\label{eq:Sn}
S^{(s)}_A = -\frac{\partial}{\partial n} S^{(s)}_n|_{n=1}\,,\quad S^{(s)}_n = \langle s | \mathcal{T}_n(z,\bar z)\mathcal{T}_{-n}(w,\bar w) | s \rangle\,,
\ee 
where 
\be
z-w= i \frac{l}{R}
\ee 
is a point in the complex plane at a distance $l$ form the origin, in a constant time slice. 

In the limit of small intervals, all the information we need about the branched worldsheet is encoded in the OPE expansion of the product of two twist fields~\cite{Calabrese:2010he,Rajabpour:2011pt,Perlmutter:2013paa}
\begin{equation}
  \label{eq:TTope}
  {\cal T}_{n}(z,\bar{z}) {\cal T}_{-n}(w,\bar w) = |z-w|^{-4\Delta}\left(1 + \sum_K (z-w)^{\Delta_K} ({\bar{z}-\bar w)^{\bar\Delta_K}} \mathcal{D}_K {\cal O}_K(0) \right)~.
\end{equation}
Here ${\cal O}_K$ is a set of quasi-primary operators living in the $n$-th product ${\cal C}^n$ of the original CFT ${\cal C}$, $\mathcal{D}_K$ is the OPE coefficient for the operator $\mathcal{O}_K$, $\Delta_K$ and $\bar\Delta_K$ are the dimensions of the holomorphic and the anti-holomorphic parts of ${\cal O}_K$, and  $\Delta=\bar\Delta=c/24 (n-1/n)$ is the conformal dimension of ${\cal T}_n$. It is important for our purposes that the operators appearing in the OPE are untwisted, i.e. they are products of operators $O^{(j)}$ defined in the original CFT on each sheet separately: ${\cal O}= O^{(1)}\otimes \ldots \otimes O^{(n)}$. By following the standard treatment used in the setup of two disjoint intervals~\cite{Calabrese:2010he,Rajabpour:2011pt,Perlmutter:2013paa} we order the contributions to the OPE~\eqref{eq:TTope} according to the number of constituents that are non-trivial ($O^{(j)}\not= 1$). So we can write
\begin{align}
  \label{eq:TTexhol}
  \mathcal{T}_n(z,\bar{z})\mathcal{T}_{-n}(w,\bar w) & = |z-w|^{-4 \Delta} \left[1 + \sum_{K,\,j} (z-w)^{\Delta_K} ({\bar{z}-\bar w) ^{\bar\Delta_K}} d^{(j)}_{K} O^{(j)}_K \right.
\\ \nonumber
& \left. +  \sum_{K,L,\,j_1\not =j_2} \! \! (z-w)^{\Delta_K+\Delta_L} ({\bar z}-\bar w)^{\bar\Delta_K+\bar\Delta_L}  d^{(j_1,j_2)}_{K L} O^{(j_1)}_K\otimes O^{(j_2)}_L + \dots\right]\,.
\end{align}
Clearly in the small $|z|$ limit we can focus on the operators with the smallest dimension. In any conformal block the operator with the smallest dimension is of course the primary operator. When only the operator on the $j$-th sheet $O^{(j)}_K$ is non-trivial, the OPE coefficients $\mathcal{D}_K$ is indicated as $d^{(j)}_{K}$. This coefficient is proportional to the 1-point function of $O^{(j)}_K$ on the $n$-th sheeted surface, which can be mapped to the complex plane by an $n$-th valued conformal map~\cite{Calabrese:2010he}; since primary operators transform homogeneously under conformal transformations, the corresponding $d^{(j)}_{K}$'s are proportional to the 1-point functions on the complex plane, that vanish for non-trivial primaries. Thus $d^{(j)}_{K}=0$ when $O^{(j)}_K$ is primary. Non-primary operators can instead have $d^{(j)}_{K}\not =0$, as it is the case for the stress energy tensor, which is the non-primary with the smallest dimension. However the states we consider are RR ground states, and in these states the stress energy tensor has vanishing vev, as it was verified in \cite{Kanitscheider:2006zf}.

Hence, the first non-trivial contribution which is of interest to us comes from the second term in (\ref{eq:TTexhol}), with non-trivial operators on two distinct copies of ${\cal C}$. We can moreover restrict the two operators to be primaries, as this will give the dominant contribution for small $|z|$. In this case the OPE coefficients will be indicated as $d^{(j_1,j_2)}_{K L}$ and have a simple general expression~\cite{Calabrese:2010he,Perlmutter:2013paa}
\begin{align}\label{eq:dj1j2}
d^{(j_1,j_2)}_{K L} & = \sum_{K',L'}(\mathcal{N}^{-1})_{KL,K'L'} \lim_{z\to\infty} |z|^{4\Delta} \langle0| {\cal T}_n(z,\bar{z}) O^{(j_1)}_{K'}\otimes O^{(j_2)}_{L'} (1) {\cal T}_{-n}(0) |0\rangle
\\ \nonumber
& = \sum_{K',L'} (\mathcal{N}^{-1})_{KL,K'L'} \mathcal{N}_{K'L'} \left(\frac{1}{2i\,n} \frac{1}{\sin \frac{\pi |j_1-j_2|}{n}}\right)^{2\Delta_{K'}} \left(\frac{1}{-2i\,n} \frac{1}{\sin \frac{\pi |j_1-j_2|}{n}}\right)^{2\bar \Delta_{K'}}\,,
\end{align}
where $\mathcal{N}_{KL,K'L'}$ is given by the vacuum two-point function of the operators in ${\cal C}^n$ (such as ${\cal O} =O^{(j_1)}_{K}\otimes O^{(j_2)}_{L}$)
\be
\mathcal{N}_{KL,K'L'} =  \langle 0| O_K(1) O_{K'}(0) |0\rangle \, \langle 0| O_L(1) O_{L'}(0) |0\rangle\,,
\ee
while the normalization $\mathcal{N}_{K'L'}$ is defined by the following correlator in ${\cal C}$
\be
\mathcal{N}_{K'L'} = \langle 0| O_{K'}(1) O_{L'}(0) |0\rangle \,.
\ee
This correlator is non-trivial only when $\Delta_{K'}=\Delta_{L'}$, which was used to simplify~\eqref{eq:dj1j2}.

Substituting (\ref{eq:TTexhol}) and (\ref{eq:dj1j2}) in (\ref{eq:Sn}), we find 
\be\label{eq:SnSmalll}
S^{(s)}_n = \frac{1}{l^{4\Delta}}\left[1+ \sum_{K,L} \Bigl(\frac{l}{2 n R}\Bigr)^{2(\Delta_K + \bar\Delta_K)} \langle O_{KL} \rangle_s \sum_{k=1}^{n-1} \frac{n-k}{\left(\sin\frac{\pi k}{n}\right)^{2(\Delta_K + \bar\Delta_K)}} \right]\,,
\ee
where 
\be
\langle O_{KL} \rangle_s \equiv \sum_{K',L'} (\mathcal{N}^{-1})_{KL,K'L'} \mathcal{N}_{K'L'} \langle s | O_{K'}  | s \rangle \langle s | O_{L'}  | s \rangle
\ee
is given in terms of the vevs of the primary operators $O_K$ in the state $|s\rangle$ computed in one copy of the original CFT $\mathcal{C}$. We have used the fact that $\mathcal{N}_{KL,K'L'}$ is non-trivial only in the subspaces with $\Delta_K=\Delta_K'$ and $\Delta_L=\Delta_L'$ to replace the $\Delta_K'$ in (\ref{eq:dj1j2}) with $\Delta_K$. The factor $n-k$ appearing in the sum over $k$ accounts for the number of terms in the sums over $j_1$ and $j_2$ with $|j_1 - j_2| = k$.

From now on, we specialize our analysis to the D1-D5 SCFT ${\cal C}_{D1D5}$ mentioned in the introduction. In particular the $SU(2)\times SU(2)$ R-symmetry of this SCFT plays an important role in our calculation. The above expression for $S_n^{(s)}$ is valid at any point in the moduli space of ${\cal C}_{D1D5}$. However, for generic primaries $O_K$, both the conformal dimensions $(\Delta_K,\bar \Delta_K)$ and the vevs $\langle O_{KL} \rangle_s $ are non-protected quantities and might depend on the couplings. This is hardly a surprise, and indeed entanglement and R\'enyi entropies do not enjoy in general any non-renormalization property.  In particular the value of $S_A^{(s)}$ derived from (\ref{eq:SnSmalll}) at the free orbifold point of ${\cal C}_{D1D5}$ does not match with the gravity result (\ref{gravityresult}). When the coupling is increased towards the regime where classical gravity is valid, most of the primary operators will get higher and higher conformal dimensions, and their contribution to (\ref{eq:SnSmalll}) will become more and more negligible. Hence to compare with gravity one should keep in (\ref{eq:SnSmalll}) only the chiral primary operators, whose dimensions are finite in the strong coupling regime. The vev of a chiral primary $O_K$ in a 1/4 BPS state is equal to the three point correlator in the vacuum of three chiral primary operators (the other two being the operators that generate the BPS state when acting on the vacuum). These correlators are known to be protected~\cite{Taylor:2007hs,Baggio:2012rr}. Hence we can compute the vevs $\langle O_K \rangle_s$ at the free point of the CFT or, holographically, from the gravity solution, and the two results should match. The holographic computation of the vevs has been done in \cite{Kanitscheider:2006zf,Kanitscheider:2007wq}. Of course to compare with gravity one should also take the limit of large central charge $c= 6 n_1 n_5 \gg 1$. As we will show below, with our conventions the 1-point functions $\langle O_K \rangle_s$ and the normalizations $\mathcal{N}_{KL}$ are proportional to $c$, and the coefficients $\mathcal{N}_{KL, K'L'}$ are proportional to $c^2$. Remembering also that the dimension of the twist fields is linear in $c$, one sees that every term in (\ref{eq:SnSmalll}) gives a contribution to the EE of order $c$. Our computation has to be contrasted with the computation of EE for two (or more) small intervals in the vacuumm\cite{Headrick:2010zt,Hartman:2013mia,Perlmutter:2013paa}: in that case one has to take the product of two (or more) copies of the OPE in (\ref{eq:TTexhol}) and evaluate their correlator in the vacuum. From~\eqref{eq:TTexhol} and~\eqref{eq:dj1j2} one can see that in this case the contribution from non-trivial primaries is of order $c^0$. Thus the EE for more than one interval in the vacuum at large $c$ is a universal quantity, which receives contributions proportional to $c$ only from the conformal block of the identity. 

In the D1-D5 SCFT, the first non-trivial chiral primaries have total dimension $\Delta_K+\bar \Delta_K =1$ and will thus contribute corrections of order $l^2$ to the EE: these are precisely the corrections expected from (\ref{gravityresult}). For operators with this conformal dimension, the sum over $k$ appearing in (\ref{eq:SnSmalll}) becomes
\be
\sum_{k=1}^{n-1} \frac{n-k}{\sin^2\frac{\pi k}{n}} = \frac{n}{2}\sum_{k=1}^{n-1} \frac{1}{\sin^2\frac{\pi k}{n}} = - 2n^2 \oint \frac{dz}{2\pi i}\frac{1}{(1-z^n) (z^2-2z + 1)}  = \frac{n(n^2-1)}{6}\,,
\ee
where in the last step we rewrote the sum as a standard anti-clockwise contour integral over $z=e^{2\pi i k/n}$ around $z=1$. 

The chiral primaries with total dimension 1 that are relevant for our purposes are:  the holomorphic and anti-holomorphic $SU(2)\times SU(2)$ currents $J^\alpha$ and $\tilde J^\alpha$ and the operators of dimension $(1/2,1/2)$ denoted as ${O}^{(1,1)}_{(1) 1}$ and ${O}^{(0,0)}_{(2)}$ in~\cite{Kanitscheider:2007wq} (actually the last two operators form quadruplets which transform as vectors of the $SO(4)$ acting on the $S^3$ coordinates).  The vevs of these operators are related with the gravity parameters $a_{\alpha\pm}$, $\mathcal{A}_{1i}$ and $f^1_{1i}$ as \cite{Kanitscheider:2007wq}
\be\label{cdef}
\langle J^\alpha \rangle_s = c_{J}\,a_{\alpha+}\,,\quad \langle \tilde J^\alpha \rangle_s =  c_{\tilde J}\,a_{\alpha-}\,,\quad \langle {O}^{(1,1)}_{(1) 1 i} \rangle_s = c_{{O}^{(1,1)}}\,\mathcal{A}_{1i}\,,\quad \langle {O}^{(0,0)}_{(2) i} \rangle_s = c_{{O}^{(0,0)}}\, f^1_{1i}\,,
\ee
where the coefficients $c$'s depend on the choice of normalization for the various operators. R-symmetry implies that the non-vanishing two-point functions are
\be
\begin{aligned}
&\langle 0| J^\alpha(1) J^\beta(0)| 0 \rangle = \mathcal{N}_{J} \,\delta^{\alpha\beta}\,,\quad \langle 0| \tilde J^\alpha(1) \tilde J^\beta(0)| 0 \rangle = \mathcal{N}_{\tilde J}\, \delta^{\alpha\beta}\,,\\
& \langle 0|  {O}^{(1,1)}_{(1) 1 i} (1) {O}^{(1,1)}_{(1) 1 j} (0) | 0 \rangle = \mathcal{N}_{{O}^{(1,1)}}\, \delta_{ij}\,,\quad \langle 0 | {O}^{(0,0)}_{(2) i} (1) {O}^{(0,0)}_{(2)j} (0) | 0 \rangle = \mathcal{N}_{{O}^{(0,0)}}\, \delta_{ij}\,.
\end{aligned}
\ee
Then the EE obtained from~\eqref{eq:Sn} and~\eqref{eq:SnSmalll} has the form
\be\label{EECFT}
S_A^{(s)} = \Bigl[2\,n_1 n_5 \log \frac{l}{R} - \frac{l^2}{12\,R^2}\, (\mathcal{N}_J^{-1}\, c_J^2\, a_+^2 + \mathcal{N}_{\tilde J}^{-1}\, c_{\tilde J}^2\, a_-^2 + \mathcal{N}^{-1}_{{O}^{(1,1)}}\,c_{{O}^{(1,1)}}^2\,\mathcal{A}_1^2+ \mathcal{N}^{-1}_{{O}^{(0,0)}}\,c_{{O}^{(0,0)}}^2\,f_1^2)\Bigr]\,,
\ee
which agrees, in structure, with the gravity result (\ref{gravityresult}). To refine the comparison and determine also the numerical coefficients, one needs to know the precise normalization of the various operators. We fix the normalizations by comparison with the particular two-charge geometry which was first considered in Section 6.4 of~\cite{Kanitscheider:2007wq}, where the corresponding state in the language of the dual CFT was also identified. An explicit representation of this state at the free orbifold point of the CFT was worked out in \cite{Giusto:2013bda}. It is straightforward to check that this state has non-trivial vevs for $J^3$, $\tilde J^3$ and ${O}^{(1,1)}_{(1) }$, and this enables us to uniquely determine the values of $c_J$, $c_{\tilde J}$, $c_{{O}^{(1,1)}}$, $\mathcal{N}_J$, $\mathcal{N}_{\tilde J}$, $\mathcal{N}_{{O}^{(1,1)}}$. The operator ${O}^{(0,0)}_{(2)}$ is of a qualitative different nature, because it involves a twist field of the orbifold CFT: we will leave the analysis of states with non-trivial vevs of this operator to a future investigation, and for the moment restrict to geometries with $f_1^2=0$.

The values of the parameters $a_{\alpha,\pm}$ and $\mathcal{A}_{1i}$ for the two-charge geometry under consideration can be read off from Eqs. (3.11) of \cite{Giusto:2013bda}, using the identifications $A\equiv -\frac{\beta+\omega}{\sqrt{2}}$, $\mathcal{A}\equiv Z_4$. After expanding these quantities for large $r$ and comparing with (\ref{eq:asymp}), one finds
the following non-trivial values
\be\label{aA}
a_{3\,+}=-a_{3\,-} = \frac{R\,a^2}{2\,\sqrt{Q_1 Q_5}}\,,\quad \mathcal{A}_{11}=\frac{R \,a \,b}{2\sqrt{Q_1 Q_5}}\,,
\ee
where the radius $R$ is related with other parameters of the geometry by
\be
R= \sqrt{\frac{Q_1 Q_5}{a^2+\frac{b^2}{2}}}\,.
\ee
The relevant CFT operators are given by\footnote{The $\chi$'s are free fermionic fields and we follow the notation of \cite{Giusto:2013bda}.}
\be
J^3 = \sum_\ell \frac{1}{2} (\chi^1_\ell  \bar \chi^1_\ell+ \chi^2_\ell  \bar \chi^2_\ell)\,,\quad \tilde J^3 = \sum_\ell \frac{1}{2} ({\tilde\chi}^1_\ell  {\tilde{\bar \chi}}^1_\ell+ {\tilde \chi}^2_\ell {\tilde {\bar \chi}}^2_\ell)\,,
\ee
\be
{O}\equiv {O}^{(1,1)}_{(1) 1 1} - i {O}^{(1,1)}_{(1) 1 2} = \frac{1}{\sqrt{2}} \sum_\ell (\bar\chi^1_\ell  {\tilde{\bar \chi}}^2_\ell- \bar\chi^2_\ell  
{\tilde{\bar \chi}}^1_\ell)\,,
\ee
where the sum over $\ell$ runs over the $n_1 n_5$ copies of the orbifold CFT. From these expressions it is immediate to compute the normalizations
\be
\mathcal{N}_J=\mathcal{N}_{\tilde J}= \mathcal{N}_{\mathcal{O}^{(1,1)}} = \frac{n_1 n_5}{2}\,.
\ee
The state dual to this geometry is
\be
|s\rangle = \sum_{k=0}^{n_1 n_5} C_k \frac{{O}^k}{k!} |n_1 n_5/2\rangle\,,
\ee
where $|n_1 n_5/2\rangle$ is the unique two-charge state with $J^3=\tilde J^3=n_1 n_5/2$, and the coefficients $C_k$ are
\be
C_k = \Bigl(\frac{R\,a}{\sqrt{Q_1 Q_5}}\Bigr)^{n_1 n_5-k} \Bigl(\frac{R\,b}{\sqrt{2\,Q_1 Q_5}}\Bigr)^k\,.
\ee
One can thus explicitly compute the vevs of the relevant operators on this state:
\be
\langle J^3 \rangle_s = \langle \tilde J^3 \rangle_s = \frac{n_1 n_5}{2} \frac{R^2\,a^2}{Q_1 Q_5}\,,\quad \langle {O} \rangle_s= n_1 n_5 \frac{R^2\,a\,b}{\sqrt{2} Q_1 Q_5}\,.
\ee
Comparing these vevs with (\ref{cdef}) and (\ref{aA}), one finds
\be
c_J= - c_{\tilde J}=n_1 n_5\,\frac{R}{\sqrt{Q_1 Q_5}}\,,\quad c_{{O}^{(1,1)}} = \sqrt{2} n_1 n_5\,\frac{R}{\sqrt{Q_1 Q_5}}\,.
\ee
Substituting in (\ref{EECFT}) one gets
\be
S_A^{(s)} = 2\,n_1 n_5\Bigl[ \log \frac{l}{R} - \frac{l^2}{12\, Q_1 Q_5}\, (a_+^2 + a_-^2 +2\mathcal{A}_1^2)\Bigr]\,,
\ee
which matches with (\ref{gravityresult}), when $f_1^2=0$.

\section{Discussion}\label{discussion}

In this paper we focused on the EE for a single interval in a $1+1$ CFT. It is well known that this quantity depends only on the CFT central charge in the simplest case~\cite{Calabrese:2004eu}, i.e. when the EE is calculated by using a density matrix obtained starting from the $SL(2,\mathbb{C})$ invariant vacuum of the CFT and tracing over the degrees of freedom outside the interval. Not surprisingly, the situation is more complicated if one starts from a generic eigenstate $|s\rangle$ of the CFT Hamiltonian. In order to discuss analytically the EE $S^{(s)}$ in these situations, we studied the short interval expansion and showed that already the first subleading term depends both on the details of the CFT and the state used to derive the density matrix. 

We focused in particular on the SCFT that is dual (in the AdS/CFT sense) to the D1-D5 system in type IIB string theory. This duality provides a precise setting where to carry out the same calculation holographically by working with explicit geometries that solve the supergravity equations. We studied in particular the simplest class of regular geometries that are $1/4$-BPS. Even in this very simple case the EE $S^{(s)}$ for a single interval depends on the details of the CFT, including the values of the various moduli. In the strongly coupled regime where supergravity is a good approximation we can compare the holographic result against the CFT expectation. In particular we showed that the holographic vevs derived in~\cite{Kanitscheider:2006zf,Kanitscheider:2007wq} are in perfect agreement with the result for the EE obtained from the generalization of the RT/HRT formula proposed in~\eqref{RTbis} that applies to 6D spacetimes asymptotic to AdS$_3\times S^3$. We thus verify that the RT/HRT holographic formalism for the computation of EE holds also in the presence of non-universal contributions.   

It is interesting to compare our results with those of~\cite{Bhattacharya:2012mi}, where the thermodynamics properties of the EE for excited states were first discussed. Since we focus on states that are not a small perturbation of the ground state, the final results are qualitatively different. For instance we have to deal in general with a non-trivial dependence on the coordinates outside the AdS space and so the natural approach is to consider the minimal area of a 4D manifold which extends in the $S^3$ directions. This also implies that the relation between the variation of the EE and the variation of the energy stored in the interval for different states does not follow the standard first law like-formula for small perturbations of the vacuum state~\cite{Bhattacharya:2012mi}. In the case analysed in this paper, all $1/4$ BPS-states have the same (zero) energy density while the EE changes. A violation of the first law-like relation for large time-dependent perturbations was also noted in \cite{Caputa:2014vaa}. 

There are of course several possible generalizations of the analysis presented here that might be interesting to pursue. We expect the generic features of the holographic calculation to be common also to higher dimensional cases, such as the $1/2$-BPS geometries of~\cite{Lin:2004nb} that are dual to states in ${\cal N}=4$ SYM. On the CFT side the EE is not captured by correlators among local operators any more, but it would still be interesting to study holographically the dependence of the EE on the particular state (geometry) considered. Another application of the approach described here is to use the EE as an observable characterising the different microstate geometries that have the same asymptotic charges. It would certainly be interesting to extend our analysis to $1/8$-BPS (three-charge) configurations and to large intervals. In the latter case the relevant manifold describing the EE extends deep inside the AdS geometry and will be sensitive to the fine details of the different microstates. However, as seen in this paper, even the first subleading term in the short interval expansion depends on the particular microstate geometry considered. So even this simple case could be used to study the relation between the EE of generic microstates and the thermal state describing the black hole with the same charges. We hope to clarify at least some of these issues in a future work.

\vspace{7mm}
 \noindent {\large \textbf{Acknowledgements} }

 \vspace{5mm} 

We would like to thank P.~Caputa, V.~Jejjala, F.~Gliozzi, L.~Martucci, K.~Skenderis, T.~Takayanagi, M.~Taylor, E.~Tonni, B.~Vercnocke for useful discussions and correspondence at several stages of this project. We also thank J. Simon for very useful comments on the first version of this article. This research is partially supported by STFC (Grant ST/J000469/1, {\it String theory, gauge theory \& duality}), by the Padova University Project CPDA119349 and by INFN.

\providecommand{\href}[2]{#2}\begingroup\raggedright\endgroup


\begin{thebibliography}{10}

\bibitem{Calabrese:2004eu}
P.~Calabrese and J.~L. Cardy, ``{Entanglement entropy and quantum field
  theory},'' {\em J.Stat.Mech.} {\bf 0406} (2004) P06002,
\href{http://arXiv.org/abs/hep-th/0405152}{{\tt hep-th/0405152}}.

\bibitem{Ryu:2006bv}
S.~Ryu and T.~Takayanagi, ``{Holographic derivation of entanglement entropy
  from AdS/CFT},'' {\em Phys.Rev.Lett.} {\bf 96} (2006) 181602,
\href{http://arXiv.org/abs/hep-th/0603001}{{\tt hep-th/0603001}}.

\bibitem{Headrick:2010zt}
M.~Headrick, ``{Entanglement Renyi entropies in holographic theories},'' {\em
  Phys.Rev.} {\bf D82} (2010) 126010,
\href{http://arXiv.org/abs/1006.0047}{{\tt 1006.0047}}.

\bibitem{Hartman:2013mia}
T.~Hartman, ``{Entanglement Entropy at Large Central Charge},''
\href{http://arXiv.org/abs/1303.6955}{{\tt 1303.6955}}.

\bibitem{Faulkner:2013yia}
T.~Faulkner, ``{The Entanglement Renyi Entropies of Disjoint Intervals in
  AdS/CFT},''
\href{http://arXiv.org/abs/1303.7221}{{\tt 1303.7221}}.

\bibitem{Lewkowycz:2013nqa}
A.~Lewkowycz and J.~Maldacena, ``{Generalized gravitational entropy},'' {\em
  JHEP} {\bf 1308} (2013) 090,
\href{http://arXiv.org/abs/1304.4926}{{\tt 1304.4926}}.

\bibitem{Caputa:2013lfa}
P.~Caputa, V.~Jejjala, and H.~Soltanpanahi, ``{Entanglement Entropy of Extremal
  BTZ},'' {\em Phys.Rev.} {\bf D89} (2014) 046006,
\href{http://arXiv.org/abs/1309.7852}{{\tt 1309.7852}}.

\bibitem{Caraglio:2008pk}
M.~Caraglio and F.~Gliozzi, ``{Entanglement Entropy and Twist Fields},'' {\em
  JHEP} {\bf 0811} (2008) 076,
\href{http://arXiv.org/abs/0808.4094}{{\tt 0808.4094}}.

\bibitem{Furukawa:2008uk}
S.~Furukawa, V.~Pasquier, and J.~Shiraishi, ``{Mutual Information and
  Compactification Radius in a c=1 Critical Phase in One Dimension},'' {\em
  Phys.Rev.Lett.} {\bf 102} (2009) 170602,
\href{http://arXiv.org/abs/0809.5113}{{\tt 0809.5113}}.

\bibitem{Calabrese:2009ez}
P.~Calabrese, J.~Cardy, and E.~Tonni, ``{Entanglement entropy of two disjoint
  intervals in conformal field theory},'' {\em J.Stat.Mech.} {\bf 0911} (2009)
  P11001,
\href{http://arXiv.org/abs/0905.2069}{{\tt 0905.2069}}.

\bibitem{Calabrese:2010he}
P.~Calabrese, J.~Cardy, and E.~Tonni, ``{Entanglement entropy of two disjoint
  intervals in conformal field theory II},'' {\em J.Stat.Mech.} {\bf 1101}
  (2011) P01021,
\href{http://arXiv.org/abs/1011.5482}{{\tt 1011.5482}}.

\bibitem{Dio:2014}
D.~Anninos and B.~Vercnocke {\em to appear} (2014).

\bibitem{Bhattacharya:2012mi}
J.~Bhattacharya, M.~Nozaki, T.~Takayanagi, and T.~Ugajin, ``{Thermodynamical
  Property of Entanglement Entropy for Excited States},'' {\em Phys.Rev.Lett.}
  {\bf 110} (2013), no.~9, 091602,
\href{http://arXiv.org/abs/1212.1164}{{\tt 1212.1164}}.

\bibitem{Nozaki:2014hna}
M.~Nozaki, T.~Numasawa, and T.~Takayanagi, ``{Quantum Entanglement of Local
  Operators in Conformal Field Theories},'' {\em Phys.Rev.Lett.} {\bf 112}
  (2014) 111602,
\href{http://arXiv.org/abs/1401.0539}{{\tt 1401.0539}}.

\bibitem{He:2014mwa}
S.~He, T.~Numasawa, T.~Takayanagi, and K.~Watanabe, ``{Quantum Dimension as
  Entanglement Entropy in 2D CFTs},''
\href{http://arXiv.org/abs/1403.0702}{{\tt 1403.0702}}.

\bibitem{Caputa:2014vaa}
  P.~Caputa, M.~Nozaki and T.~Takayanagi,
  arXiv:1405.5946 [hep-th].

\bibitem{Strominger:1996sh}
A.~Strominger and C.~Vafa, ``{Microscopic origin of the Bekenstein-Hawking
  entropy},'' {\em Phys.Lett.} {\bf B379} (1996) 99--104,
  \href{http://arXiv.org/abs/hep-th/9601029}{{\tt hep-th/9601029}}.

\bibitem{Skenderis:2008qn}
K.~Skenderis and M.~Taylor, ``{The fuzzball proposal for black holes},'' {\em
  Phys. Rept.} {\bf 467} (2008) 117--171,
\href{http://arXiv.org/abs/0804.0552}{{\tt 0804.0552}}.

\bibitem{Balasubramanian:2008da}
V.~Balasubramanian, J.~de~Boer, S.~El-Showk, and I.~Messamah, ``{Black Holes as
  Effective Geometries},'' {\em Class.Quant.Grav.} {\bf 25} (2008) 214004,
  \href{http://arXiv.org/abs/0811.0263}{{\tt 0811.0263}}.

\bibitem{Chowdhury:2010ct}
B.~D. Chowdhury and A.~Virmani, ``{Modave Lectures on Fuzzballs and Emission
  from the D1-D5 System},''
\href{http://arXiv.org/abs/1001.1444}{{\tt 1001.1444}}.

\bibitem{Mathur:2012zp}
S.~D. Mathur, ``{Black Holes and Beyond},'' {\em Annals Phys.} {\bf 327} (2012)
  2760--2793,
\href{http://arXiv.org/abs/1205.0776}{{\tt 1205.0776}}.

\bibitem{Mathur:2012dxa}
S.~D. Mathur, ``{Black holes and holography},'' {\em J.Phys.Conf.Ser.} {\bf
  405} (2012) 012005,
\href{http://arXiv.org/abs/1207.5431}{{\tt 1207.5431}}.

\bibitem{Bena:2013dka}
I.~Bena and N.~P. Warner, ``{Resolving the Structure of Black Holes:
  Philosophizing with a Hammer},''
\href{http://arXiv.org/abs/1311.4538}{{\tt 1311.4538}}.

\bibitem{Mathur:2005zp}
S.~D. Mathur, ``{The fuzzball proposal for black holes: An elementary
  review},'' {\em Fortsch. Phys.} {\bf 53} (2005) 793--827,
\href{http://arXiv.org/abs/hep-th/0502050}{{\tt hep-th/0502050}}.

\bibitem{Lunin:2001jy}
O.~Lunin and S.~D. Mathur, ``{AdS/CFT duality and the black hole information
  paradox},'' {\em Nucl. Phys.} {\bf B623} (2002) 342--394,
\href{http://arXiv.org/abs/hep-th/0109154}{{\tt hep-th/0109154}}.

\bibitem{Lunin:2002iz}
O.~Lunin, J.~M. Maldacena, and L.~Maoz, ``{Gravity solutions for the D1-D5
  system with angular momentum},''
\href{http://arXiv.org/abs/hep-th/0212210}{{\tt hep-th/0212210}}.

\bibitem{Kanitscheider:2007wq}
I.~Kanitscheider, K.~Skenderis, and M.~Taylor, ``{Fuzzballs with internal
  excitations},'' {\em JHEP} {\bf 06} (2007) 056,
\href{http://arXiv.org/abs/0704.0690}{{\tt 0704.0690}}.

\bibitem{Hubeny:2007xt}
V.~E. Hubeny, M.~Rangamani, and T.~Takayanagi, ``{A Covariant holographic
  entanglement entropy proposal},'' {\em JHEP} {\bf 0707} (2007) 062,
\href{http://arXiv.org/abs/0705.0016}{{\tt 0705.0016}}.

\bibitem{Dong:2013qoa}
X.~Dong, ``{Holographic Entanglement Entropy for General Higher Derivative
  Gravity},'' {\em JHEP} {\bf 1401} (2014) 044,
\href{http://arXiv.org/abs/1310.5713}{{\tt 1310.5713}}.

\bibitem{Camps:2013zua}
J.~Camps, ``{Generalized entropy and higher derivative Gravity},'' {\em JHEP}
  {\bf 1403} (2014) 070,
\href{http://arXiv.org/abs/1310.6659}{{\tt 1310.6659}}.

\bibitem{Bhattacharyya:2014yga}
A.~Bhattacharyya and M.~Sharma, ``{On entanglement entropy functionals in
  higher derivative gravity theories},''
\href{http://arXiv.org/abs/1405.3511}{{\tt 1405.3511}}.

\bibitem{Avery:2010qw}
S.~G. Avery, ``{Using the D1D5 CFT to Understand Black Holes},''
\href{http://arXiv.org/abs/1012.0072}{{\tt 1012.0072}}.

\bibitem{Kim:1985ez}
H.~Kim, L.~Romans, and P.~van Nieuwenhuizen, ``{The Mass Spectrum of Chiral N=2
  D=10 Supergravity on S**5},'' {\em Phys.Rev.} {\bf D32} (1985)
389.

\bibitem{Skenderis:2006uy}
K.~Skenderis and M.~Taylor, ``{Kaluza-Klein holography},'' {\em JHEP} {\bf
  0605} (2006) 057,
\href{http://arXiv.org/abs/hep-th/0603016}{{\tt hep-th/0603016}}.

\bibitem{Giusto:2013bda}
S.~Giusto and R.~Russo, ``{Superdescendants of the D1D5 CFT and their dual
  3-charge geometries},'' {\em JHEP} {\bf 1403} (2014) 007,
\href{http://arXiv.org/abs/1311.5536}{{\tt 1311.5536}}.

\bibitem{Jevicki:1998bm}
A.~Jevicki, M.~Mihailescu, and S.~Ramgoolam, ``{Gravity from CFT on S**N(X):
  Symmetries and interactions},'' {\em Nucl.Phys.} {\bf B577} (2000) 47--72,
\href{http://arXiv.org/abs/hep-th/9907144}{{\tt hep-th/9907144}}.

\bibitem{Lunin:2000yv}
O.~Lunin and S.~D. Mathur, ``{Correlation functions for M(N)/S(N) orbifolds},''
  {\em Commun. Math. Phys.} {\bf 219} (2001) 399--442,
\href{http://arXiv.org/abs/hep-th/0006196}{{\tt hep-th/0006196}}.

\bibitem{Lunin:2001pw}
O.~Lunin and S.~D. Mathur, ``{Three-point functions for M(N)/S(N) orbifolds
  with N = 4 supersymmetry},'' {\em Commun. Math. Phys.} {\bf 227} (2002)
  385--419,
\href{http://arXiv.org/abs/hep-th/0103169}{{\tt hep-th/0103169}}.

\bibitem{Rajabpour:2011pt}
M.~Rajabpour and F.~Gliozzi, ``{Entanglement Entropy of Two Disjoint Intervals
  from Fusion Algebra of Twist Fields},'' {\em J.Stat.Mech.} {\bf 1202} (2012)
  P02016,
\href{http://arXiv.org/abs/1112.1225}{{\tt 1112.1225}}.

\bibitem{Perlmutter:2013paa}
E.~Perlmutter, ``{Comments on Renyi entropy in AdS$_3$/CFT$_2$},'' {\em JHEP}
  {\bf 1405} (2014) 052,
\href{http://arXiv.org/abs/1312.5740}{{\tt 1312.5740}}.

\bibitem{Kanitscheider:2006zf}
I.~Kanitscheider, K.~Skenderis, and M.~Taylor, ``{Holographic anatomy of
  fuzzballs},'' {\em JHEP} {\bf 0704} (2007) 023,
\href{http://arXiv.org/abs/hep-th/0611171}{{\tt hep-th/0611171}}.

\bibitem{Taylor:2007hs}
M.~Taylor, ``{Matching of correlators in AdS(3) / CFT(2)},'' {\em JHEP} {\bf
  0806} (2008) 010,
\href{http://arXiv.org/abs/0709.1838}{{\tt 0709.1838}}.

\bibitem{Baggio:2012rr}
M.~Baggio, J.~de~Boer, and K.~Papadodimas, ``{A non-renormalization theorem for
  chiral primary 3-point functions},'' {\em JHEP} {\bf 1207} (2012) 137,
\href{http://arXiv.org/abs/1203.1036}{{\tt 1203.1036}}.

\bibitem{Lin:2004nb}
H.~Lin, O.~Lunin, and J.~M. Maldacena, ``{Bubbling AdS space and 1/2 BPS
  geometries},'' {\em JHEP} {\bf 10} (2004) 025,
\href{http://arXiv.org/abs/hep-th/0409174}{{\tt hep-th/0409174}}.

\end{thebibliography}

\end{document}